\documentstyle[12pt]{article}

\parskip=\medskipamount
\begin{document}

\begin{titlepage}

\begin{center}
{ \large \bf Intensity-dependent pion-nucleon coupling in  
multipion production processes } \\
\vspace{2.cm}
{ \normalsize \sc  M. Martinis and V. Mikuta-Martinis\footnote
{e-mail address: martinis@thphys.irb.hr \\
e-mail address: mikutama@thphys.irb.hr}} \\
\vspace{0.5cm}
Department of  Physics, \\
Theory Division, \\
Rudjer Bo\v skovi\' c Institute, P.O.B. 1016, \\
10001 Zagreb,CROATIA \\
\vspace{2cm}
{ \large \bf Abstract}
\end{center}
\vspace{0.5cm}
\baselineskip=24pt
We propose an intensity-dependent pion-nucleon coupling Hamiltonian
within a unitary multiparticle-production  model of the 
Auerbach-Avin-Blankenbecler-Sugar (AABS) type in which 
the  pion field is represented
by the thermal-density matrix. Using this Hamiltonian, we explain 
the appearance of the negative-binomial (NB) distribution for pions and 
the well-known empirical relation, the so-called Wr\' oblewski relation, 
in which the  dispersion $D$ of the pion-multiplicity distribution
is linearly related to the average multiplicity $<n>$ :
$D = A<n> + B $, with the coefficient $A < 1$.  
 The Hamiltonian of our  model is expressed 
linearly in terms of the generators of the $SU(1,1)$ group.
We also find the generating function for the pion field, which
reduces to  the generating function of the NB distribution
in the limit $T \to 0$.

PACS number(s): 12.40.Ee

\end{titlepage}

\setcounter{page}{1}
\newpage
\baselineskip=24pt
\vspace{1.5cm}
\section{Introduction}

During the last years a considerable amount of experimental information
has been accumulated on multiplicity distributions of charged
particles produced in $pp$ and $p\bar{p}$ collisions in the centre- 
of-mass energy range from $10 GeV$  to $1800 GeV$ .
Measurements in the regime of several hundred $GeV$ [1] have shown
the violation of the Koba-Nielsen-Olesen (KNO)
scaling [2], which was previously observed in the
 ISR c.m.energy range from 11 to 63 $GeV$  [3].
The violation of the KNO scaling is characterized by an
enhancement of high-multiplicity events leading to a 
broadening of the multiplicity distribution with energy.

The shape of the multiplicity distribution may be described
by either its \\
C moments, $C_{q} = \langle n^{q}\rangle / 
\langle n\rangle ^{q}$, or by  its central moments (higher-order dispersions), 
$D_{q} = \langle (n - \langle n\rangle )^{q} 
\rangle ^{1/q},  q = 2,3 \ldots $. 
The exact KNO scaling implies that all $C_{q}$ 
moments are energy independent. The $C$ moments appear to be energy 
independent only at energies below $100 GeV$.
It can  also been shown [4] that the KNO scaling  leads to a
generalized Wr\' oblewski relation  [5]   
\begin{equation}
D_{q} = A_{q}\langle n\rangle  - B_{q},
\end{equation}
with the energy-independent coefficients $A_{q}$ and $B_{q}$.
The $pp$ and $p\bar{p}$ inelastic data below $100 GeV$  also
show a linear dependence of the dispersion on
the average number of charged particles, but with the coefficients
$A_{q}$ and $B_{q}$ that are approximately equal within errors.

The fact that the dispersion of the multiplicity distribution
grows linearly with $\langle n\rangle $ implies that the
elementary Poisson distribution resulting from the independent
emission of particles is ruled out.

The total multiplicity distribution $P_{n}$ of charged particles
for a wide range of energies $(22 - 900 GeV)$
is found to be well described by a negative-binomial (NB)
distribution [1,6] that belongs 
to a large class of compound Poisson distributions [7]. 
It is a two-step process [8] with two free parameters:
the average number of charged particles  $\langle n\rangle $
and the parameter $k$ which affects the shape (width) of the
distribution. The parameter $k$ is also related to the
dispersion $D = D_{2}$ by the relation
\begin{equation}
(\frac{D}{\langle n\rangle })^{2}  = \frac{1}{k} + \frac{1}{\langle n\rangle },
\end{equation}
so that the observed broadening of the normalized multiplicity
distribution with increasing energy implies a decrease of the
parameter $k$ with energy. The KNO scaling requires $k$ to be
constant.

Although the NB distribution
gives information on the structure of correlation functions in 
multiparticle production,  the question still remains 
whether its clan-structure interpretation  is
simply a new parametrization of the data or  has deeper physical
insight [9]. Measurements  of multiplicity distributions in $p\bar{p}$
collisions at $TeV$ energies [10] have recently shown that their shape 
is clearly different from that of the NB distribution.
The distributions display the so-called medium-multiplicity
"shoulder", with a shape qualitatively similar to that 
of the UA5 $900 GeV$  and UA1 distributions [11]. A satisfactory
explanation of this effect is still lacking [12].

In this paper we propose another approach to multiplicity
distributions based on a unitary eikonal model with a 
pion-field thermal-density operator given in terms of an effective 
intensity-dependent pion-nucleon coupling Hamiltonian. We assume that  
the system of produced hadronic matter  behaves as a 
hadron gas  in thermodynamical
equilibrium at the temperature $T$ before the hadrons themselves 
decouple (freezing-out) and decay producing observable particles in 
the detector.

The paper is organized as follows.
In Sect.2  we explain the basic ideas of our unitary eikonal model 
with a  pion-field thermal-density operator.  
A discussion of the Wr\' oblewski relation and the NB distribution 
are presented in Sect. 3.
Finally, in Sect. 4 we draw conclusions and make remarks on a possible
extension of the model to include  two-pion correlations in the 
effective pion-nucleon Hamiltonian.

\newpage
\section{Description of the model}

At present accelerator energies the number of secondary
particles (mostly pions) produced in hadron-hadron collisions
is large enough, so that the statistical approach to particle
production becomes reasonable . Most of the properties of
pions produced in  high-energy hadron-hadron collisions
can be expressed simply in terms of a pion-field density
operator. We neglect difficulties associated with isospin
and only consider the production of  isoscalar "pions".
In high-energy collisions, most of the pions
are produced in the central region. In this region the energy - momentum
conservation has a minor effect if the transverse momenta of the
pions are limited by the dynamics.

{\bf A. The AABS model}

In a class of unitary eikonal models  (AABS models) [13] 
formulated a long time ago, the incident hadrons propagate
through the interaction region, without making significant
changes in their longitudinal momenta (leading-particle effect).
Only the part $W = K\sqrt{s}$ of the total c.m. energy $\sqrt{s}$  in
every concrete event is avaliable for particle production, where
$K$ is the inelasticity  $0\le K \le 1$ .

In the AABS type of models, the scattering operator $\hat{S}$ is
diagonal in the rapidity difference $Y = ln(s/m^{2})$ 
and in the relative impact
parameter $\vec{B}$ of the two incident hadrons. 
The initial-state vector for the pion field
is  $\hat{S}(Y,\vec{B})\mid 0 >$, where the vacuum state $\mid 0 >$
for pions is in fact a state containing two incident hadrons. 

The $n$-pion production amplitude for $n\ge 1$ is  given  by 
\begin{equation}
iT_{n}(Y,\vec{B}; k_{1}\ldots k_{n}) = 2s\langle k_{1}\ldots k_{n}
\mid \hat{S}(Y,\vec{B})\mid 0\rangle .
\end{equation}
The square of the $n$-pion production amplitude can be written in the form
\begin{equation}
\mid T_{n}(Y,\vec{B}; k_{1}\ldots k_{n})\mid ^{2} =
4s^{2}Tr\{\rho (Y,\vec{B})\mid k_{1}\ldots k_{n}\rangle \langle k_{1}
\ldots k_{n}\mid \},
\end{equation}
where the pion-density operator $\rho (Y,\vec{B})$ is defined as
\begin{equation}
\rho (Y,\vec{B}) = \hat{S}(Y,\vec{B})\mid 0\rangle \langle 0\mid
\hat{S}^{\dag }(Y,\vec{B}).
\end{equation}
The square of the elastic scattering amplitude is then  the
matrix element of $\rho (Y,\vec{B})$ between the states with no
pions, i.e. $\langle 0\mid \rho (Y,\vec{B})\mid 0\rangle $.

In terms of the pion-number operator
\begin{equation}
\hat{N} = \sum_{k} a^{\dag }_{k}a_{k} = \sum_{k} \hat{N}_{k} , \,\,
k \equiv (\omega _{k},\vec{k}),
\end{equation}
the square of the $S$-matrix element when no pions are emitted can 
also be written in the form
\begin{equation}
\mid \langle 0\mid \hat{S}(Y,\vec{B})\mid 0\rangle \mid ^{2} = 
Tr\{ \rho (Y,\vec{B}) : e^{-\hat{N}} :\} = e^{- \Omega (Y,\vec{B})}.
\end{equation}
Here $: \, :$ indicates the operation of normal ordering and
$\Omega (Y,\vec{B})$ is the usual eikonal function (or the
opacity function) of the geometrical model [14]. The connection
with the inelastic cross section and the exclusive cross section
for production of $n$ pions is then
\begin{equation}
\sigma _{inel}(Y,\vec{B}) = 1 - e^{- \Omega (Y,\vec{B})},
\end{equation}
and for $n\geq 1$, it is
\begin{equation}
\sigma _{n}(Y,\vec{B}) = Tr\{ \rho (Y,\vec{B}) : \frac{\hat{N}^{n}}{n!}
e^{- \hat{N}} :\}.
\end{equation}
In terms of a normalized pion-multiplicity distribution at each impact
parameter, $P_{n}(Y,\vec{B}) = \sigma _{n}
(Y,\vec{B})/ \sigma _{inel}(Y,\vec{B})$, the observed complete 
multiplicity distribution $P_{n}(Y)$ is obtained by summing $P_{n}(Y,\vec{B})$
over all impact parameters $\vec{B}$ with the weight function
$Q(Y,\vec{B}) = \sigma_{inel}(Y,\vec{B}) / \sigma_{inel}(Y)$, i.e. 
\begin{equation}
P_{n}(Y) = \int d^{2}B Q(Y,\vec{B}) P_{n}(Y,\vec{B}).
\end{equation}

The first-order moment of $P_{n}(Y)$ gives the average  multiplicity
\begin{equation}
\langle n\rangle = \sum n P_{n}(Y) = \int d^{2}B Q(Y,\vec{B})
\bar{n}(Y,\vec{B}).
\end{equation}
The higher-order moments of $P_{n}(Y)$ give information on the 
dynamical fluctuations from $\langle n\rangle $ and also on the
multiparticle correlations. All these higher-order moments can be
obtained by differentiation from the pion-generating function
\begin{equation}
G(z) = \sum z^{n}P_{n}(Y) = \int d^{2}B Q(Y,\vec{B})G(Y,\vec{B};z),
\end{equation}
where
\begin{equation}
G(Y,\vec{B};z) = Tr\{\rho (Y,\vec{B})z^{\hat{N}}\}
\end{equation}
is the pion-generating function in $B$-space .
Thus the normalized factorial moments $F_{q}$ are
\begin{equation}
F_{q} = \frac{\langle n(n-1)\ldots (n-q+1)\rangle}{\langle n\rangle ^{q}} = 
\langle n\rangle ^{- q}\frac{d^{q}G(1)}{dz^{q}} 
\end{equation}
and the normalized cumulant moments $K_{q}$ are
\begin{equation}
K_{q} = \langle n\rangle ^{- q}\frac{d^{q}lnG(1)}{dz^{q}}.
\end{equation}
These moments are related to each other by the formula  
\begin{equation}
F_{q} = \sum_{l=0}^{q-1} {q-1 \choose l}K_{q-l}F_{l}.
\end{equation}
For the Poisson distribution, all the normalized factorial moments are
identically equal to unity and all  cumulants vanish for $q > 1$.

We shall be concern mostly with the $q =2$ moments, which are directly
related to the dispersion $D$:
\begin{equation}
F_{2} = K_{2} + 1 = (\frac{D}{\langle n\rangle })^{2} + 
1 - \frac{1}{\langle n\rangle }.
\end{equation}

{\bf B. Thermal-density operator for the pion field}

The operator $\mid 0\rangle \langle 0\mid $ appearing in the definition
of $\rho (Y,\vec{B})$ represents the density operator $\rho (vac)$ for
the pion-field vacuum state.
 
The density operator for a
pion field in thermal equilibrium at the temperature $T$ is  
\begin{equation}
\rho _{T} = \frac{1}{Z} e^{- \beta H_{0}} , \,\, \beta = \frac{1}{k_{B}T},
\end{equation}
where
\begin{eqnarray}
H_{0} & = & \sum_{k} \omega _{k}(a^{\dag }_{k}a_{k} + \lambda ), \\
  lnZ & = & - \beta \lambda \sum_{k} \omega _{k} -  
\sum_{k} ln(1 - e^{-\beta \omega _{k}}). \nonumber
\end{eqnarray}
The quantity $\lambda \sum_{k}\omega _{k} = 
\langle 0\mid H_{0}\mid 0\rangle - 
\sum_{k}\omega _{k}\langle 0\mid \hat{N}_{k}\mid 0\rangle  $ 
represents the lowest possible
energy of the pion system in the leading - particle environment.
The " zero-point
energy " corresponds to $\lambda = \frac{1}{2}$.  
If the energies $\omega _{k} = \sqrt{\vec{k}^{2} + m_{\pi }^{2}}$ 
of the pion gas in the volume $V$ are
closely spaced, the summation over $k$ is replaced by an integral:
$\sum_{k} \to V\int d^{3}k/2\omega _{k}$. \\
Note that $\rho (vac) = \rho _{T=0}$.
The mean number of thermal (chaotic) pions is
\begin{eqnarray}
\bar{n}_{T} & = & \sum_{k} \frac{1}{e^{\beta \omega _{k}} - 1}  \\  \nonumber
& = & \sum_{k}\bar{n}_{Tk}.   
\end{eqnarray}

Owing to the interaction of pions with the nucleon field the density 
operator  $\rho _{T}$ is transformed by means of the unitary $S$-matrix 
into
\begin{eqnarray}
\rho _{T}(Y,\vec{B}) & = & \hat{S}(Y,\vec{B}) \rho _{T} 
\hat{S^{\dag}}(Y,\vec{B}) \\  \nonumber
                     & = & \frac{1}{Z}e^{- \beta H(Y,\vec{B})},
\end{eqnarray}
where 
\begin{equation}
H(Y,\vec{B}) = \hat{S}(Y,\vec{B})H_{0}\hat{S^{\dag}}(Y,\vec{B})
\end{equation}
is regarded as an effective Hamiltonian describing the pion system
in the interaction with the leading - particle system.
 
Now, taking into account an old observation of Golab-Meyer and Ruijgrok [15]
that the Wr\' oblewski relation can be satisfied for all energies if
the square of the pion-nucleon coupling constant increases linearly
with the mean number of pions $\langle n\rangle $, we propose the 
following form of the effective pion-nucleon 
coupling Hamiltonian:
\begin{eqnarray}
H(Y,\vec{B}) & = & \sum_{k} [ \epsilon _{k}(Y,\vec{B})(N_{k} + \lambda ) + 
g_{k}(Y,\vec{B})( a_{k}\sqrt{N_{k} + 2\lambda  - 1} + h.c.) ] \\ 
             & = & \sum_{k} H_{k}(Y,\vec{B}),  \nonumber
\end{eqnarray}
where $\epsilon _{k}^{2}(Y,\vec{B}) = \omega _{k}^{2} + 4g_{k}^{2}(Y,\vec{B})$.
The interaction part of the Hamiltonian $H_{k}$ for the k mode is no longer
linear in the pion-field variables $a_{k}$ and represents an 
intensity-dependent coupling [16]. It is also easy to see that the operators
\begin{eqnarray}
K_{0}(k) & = & N_{k} + \lambda  , \\  \nonumber
K_{-}(k) & = & a_{k}\sqrt{N_{k} + 2\lambda - 1} ,  \\
K_{+}(k) & = & \sqrt{N_{k} + 2\lambda - 1}\,a_{k}^{\dag }  \nonumber
\end{eqnarray}
form the standard Holstein-Primakoff [17] realizations of the $su(1,1)$ Lie 
algebra, the Casimir operator of which is
\begin{equation}
\hat{C}_{k} = K_{0}^{2}(k) - \frac{1}{2}[K_{+}(k)K_{-}(k) + K_{-}(k)K_{+}(k)] =
\lambda (\lambda  - 1)\hat{I}_{k}.
\end{equation} 
The Hamiltonian  $H_{k}(Y,\vec{B})\equiv  H_{k}$ is thus a linear 
combination of the generators of the  $SU(1,1)$ group: 
\begin{equation}
H_{k} = \epsilon _{k} K_{0}(k) + g_{k}[K_{+}(k) + K_{-}(k)].
\end{equation}
The corresponding S-matrix which diagonalizes the Hamiltonian $H(Y,\vec{B})$
is 
\begin{equation}
\hat{S}(Y,\vec{B}) = \prod_{k} \hat{S}_{k}(Y,\vec{B}),
\end{equation}
where
\begin{equation}
\hat{S}_{k}(Y,\vec{B}) = exp\{ - \theta _{k}(Y,\vec{B})[K_{+}(k) - 
K_{-}(k)]\},
\end{equation}
with
\begin{equation}
th\,\theta _{k}(Y,\vec{B}) = \frac{2g_{k}(Y,\vec{B})}{\epsilon _{k}(Y,\vec{B})}.
\end{equation}
Since the dependence on the variables $Y,\vec{B}$ is contained only
in the hyperbolic angle $\theta _{k}(Y,\vec{B})$, we shall from now on
assume this dependence whenever we write $\theta _{k}$.

It is  easy to see that the initial-state vector for the pion field,
$\hat{S}(Y,\vec{B})\mid 0\rangle $, factorizes in the $k$-space as
\begin{equation}
\hat{S}(Y,\vec{B})\mid 0\rangle  = \prod _{k}(\hat{S}_{k}(Y,\vec{B})
\mid 0_{k}\rangle ),
\end{equation}
with  
\begin{eqnarray}
\hat{S}_{k}(Y,\vec{B})\mid 0_{k}\rangle & = & (1 - th^{2}\theta _{k})^{\lambda }
\sum _{n_{k}}(- th\theta _{k})^{n_{k}}\big(\frac{\Gamma (n_{k} + 2\lambda )}
{n_{k}!\Gamma (2\lambda )}\big)^{1/2}\mid n_{k}\rangle  \\
& = & \mid \theta _{k}\rangle ,  \nonumber
\end{eqnarray} 
where $\mid n_{k}\rangle  = (n_{k}!)^{- 1/2}(a^{\dag }_{k})^{n_{k}}
\mid 0_{k}\rangle $.
In the same way we find that the pion thermal-density operator 
$\rho _{T}(Y,\vec{B})$ is also factorized as 
\begin{equation}
\rho _{T}(Y,\vec{B}) = \prod _{k}\rho _{T}(\theta _{k}),
\end{equation}
with
\begin{equation}
\rho _{T}(\theta _{k}) = \frac{1}{Z_{k}}\sum _{n_{k}} e^{- \beta \omega _{k}
(n_{k} + \lambda )}\mid n_{k},\theta _{k}\rangle \langle n_{k},\theta _{k}\mid ,
\end{equation}
where $\mid n_{k},\theta _{k}\rangle  = \hat{S}_{k}(Y,\vec{B})\mid n_{k}\rangle $.
The states $\mid n_{k},\theta _{k}\rangle $ form a  complete orthonormal
set of eigenvectors of the $k$-mode
Hamiltonian $H_{k}$, i.e.
\begin{eqnarray}
H_{k}\mid n_{k},\theta _{k}\rangle  & = & \omega _{k}(n_{k} + \lambda )
\mid n_{k},\theta _{k}\rangle ,\\ 
\sum _{n_{k}}\mid n_{k},\theta _{k}\rangle  
\langle n_{k},\theta _{k}\mid  & = &  I, \\ 
\langle n_{k},\theta _{k}\mid m_{k},\theta _{k}\rangle & = &  \delta _{n,m}.
\end{eqnarray}
 
\newpage 
\section{Pion-generating function and its moments }

The average multiplicity $\bar{n}_{T}(Y,\vec{B})$, the dispersion
$d_{T}^{2}(Y,\vec{B})$ and all  higher-order moments at the
temperature $T$ in $B$ space,  
\begin{equation}
\bar{n^{q}_{T}}(Y,\vec{B}) = Tr\{\rho _{T}(Y,\vec{B})\hat{N}^{q}\},\, 
q = 1,2,\ldots 
\end{equation} 
can be obtained by differentiation from the pion-generating function  
\begin{equation}
G_{T}(Y,\vec{B};z) = \prod _{k}G_{T}(\theta _{k};z),
\end{equation}
where 
\begin{equation}
G_{T}(\theta _{k};z) = Tr\{\rho _{T}(\theta _{k})z^{\hat{N}_{k}}\}.
\end{equation}

After performing a certain amount of straightforward 
algebraic manipulations, we find
the following expression for the pion-generating function 
$G_{T}(\theta _{k};z)$:
\begin{equation}
G_{T}(\theta _{k};z) = G_{0}(\theta _{k};z)(1 - e^{- \beta \omega _{k}})
2^{2\lambda - 1}R_{k}^{-1}(1 + y_{k}  + R_{k})^{1 - 2\lambda }, 
\end{equation}
where 
\begin{eqnarray}
R_{k} & = & \sqrt{1 - 2x_{k}y_{k} + y_{k}^{2}},  \\   \nonumber
x_{k} & = & \frac{z + (1-z)^{2}sh^{2}(\theta _{k})ch^{2}(\theta _{k})}
{z - (1-z)^{2}sh^{2}(\theta _{k})ch^{2}(\theta _{k})},  \\
y_{k} & = & e^{-\beta \omega _{k}}\frac{z - (1-z)sh^{2}(\theta _{k})}
{1 + (1-z)^{2}sh^{2}(\theta _{k})},  \nonumber 
\end{eqnarray}
and $G_{0}(\theta _{k};z)$ denotes the pion-generating function 
at the temperature $T = 0$:
\begin{equation}
G_{0}(\theta _{k};z) = [ 1 + (1-z)sh^{2}(\theta _{k})]^{- 2\lambda }.
\end{equation}
We observe that $G_{0}$ is exactly the generating function of the NB  
distribution with a constant shape parameter  $2\lambda $, and the 
average number of k-mode pions is equal to 
\begin{equation}
\bar{n}(\theta _{k}) = 2\lambda sh^{2}(\theta _{k}).
\end{equation} 
The vacuum value of the $k$-mode thermal-density operator 
$\rho _{T}(\theta _{k})$ is used to obtain the
$k$-mode thermal eikonal function $\Omega _{T}(\theta _{k})$
\begin{eqnarray}
\langle 0_{k}\mid \rho _{T}(\theta _{k})\mid 0_{k}\rangle  & = &
e^{-\Omega _{T}(\theta _{k})}  \\  \nonumber
& = & (1 - e^{- \beta \omega _{k}})G_{0}(\theta _{k};e^{- \beta \omega _{k}}).
\end{eqnarray}
The total eikonal function is 
$\Omega _{T}(Y,\vec{B}) = \sum_{k}\Omega _{T}(\theta _{k}) $.

For the $k$-mode pion field in $B$ space , we 
find the following average number and the dispersion: 
\begin{eqnarray}
\bar{n}_{T}(\theta _{k}) & = & \bar{n}(\theta _{k}) + \bar{n}_{Tk} + 
\frac{1}{\lambda }\bar{n}(\theta _{k})\bar{n}_{Tk},  \\  \nonumber
d_{T}^{2}(\theta _{k}) & = & d_{Tk}^{2} + d^{2}(\theta _{k})
[ 1 + \frac{2\lambda - 3}{\lambda }\bar{n}_{Tk} + \frac{4}{\lambda }
\bar{n}_{Tk}^{2}],
\end{eqnarray} 
where
\begin{eqnarray}
d_{Tk}^{2} & = & \bar{n}_{Tk}^{2} + \bar{n}_{Tk},  \nonumber  \\
d^{2}(\theta _{k}) & = & \frac{1}{2\lambda }\bar{n}^{2}(\theta _{k}) +
\bar{n}(\theta _{k}).
\end{eqnarray}

Two limiting cases are of interest, namely $T \to 0$ and $T \to\infty $.\\ 
For $T \to 0$ , we find
\begin{equation}
\frac{d^{2}(\theta _{k})}{\bar{n}^{2}(\theta _{k})} = 
\frac{1}{2\lambda } + \frac{1}{\bar{n}(\theta _{k})}, 
\end{equation}
as it is to be expected from the  NB distribution . 
However,  the interpretation of this result is quite different. 
In our case, the parameter $\lambda $ is connected with 
the vacuum expectation value of the effective Hamiltonian $H(Y,\vec{B})$.
It has nothing to do
with either the number of pion sources or the number of clans.
Since $SU(1,1)$ is a dynamical symmetry group of our effective Hamiltonian, 
the parameter $\lambda $ also labels the positive discrete class of its 
unitary irreducible representations.
It is important
to observe that pions in the $k$ mode are distributed according to
the NB distribution with a constant shape parameter $2\lambda $.
The Wr\' oblewski relation
\begin{equation}
d(\theta _{k}) = A\bar{n}(\theta _{k}) + B
\end{equation}
obtained from (47) gives the energy-independent coefficients $A = (2\lambda )^{- 1/2}$ and $B = (\lambda /2)^{1/2}$. If $\lambda  > 1/2$, we have $A < 1$.

 The contribution from all the $k$ modes in $B$ space gives
\begin{equation}
\frac{d^{2}(Y,\vec{B})}{\bar{n}^{2}(Y,\vec{B})} = \frac{1}{2\lambda }
\sum_{k}p^{2}(\theta _{k})  +  \frac{1}{\bar{n}(Y,\vec{B})},
\end{equation}
where $p(\theta _{k}) = \bar{n}(\theta _{k})/\bar{n}(Y,\vec{B})$. In
this case, the coefficient $A$ in the Wr\' oblewski relation becomes
energy and $B$ dependent and is of the form
\begin{equation}
A(Y,\vec{B}) = \big[ \frac{1}{2\lambda }\sum_{k}p^{2}(\theta _{k})\big]^{1/2}.
\end{equation}
Since $\sum_{k}p(\theta _{k}) = 1$ and all $p(\theta _{k})$ are positive
functions of $\theta _{k}$, the sum $\sum_{k}p^{2}(\theta _{k})$ is always
smaller than unity. Therefore, $A(Y,\vec{B}) < 1$ if $\lambda  > 1/2$.

Finally, the summation over all impact parameters gives 
\begin{equation}
\big(\frac{D}{\langle n\rangle }\big)^{2} = \int d^{2}B Q(Y,\vec{B})
[(A^{2}(Y,\vec{B}) + 1)\big(\frac{\bar{n}(Y,\vec{B})}
{\langle n\rangle }\big)^{2} - 1] + \frac{1}{\langle n\rangle }.
\end{equation}
This expression, when combined with our preceding analysis, suggests that
the coefficient $A$ in the Wr\' oblewski relation  should be energy
dependent and smaller than unity.

For the temperature $T$ going to infinity we obtain 
\begin{eqnarray}
\frac{d_{T}^{2}(\theta _{k})}{\bar{n}_{Tk}^{2}(\theta _{k})}
\Big|_{T\to\infty} & = & 2 - (1 + \frac{\bar{n}(\theta _{k})}
{\lambda })^{- 2}   \\   \nonumber
& = &  1 + th^{2}(2\theta _{k}).
\end{eqnarray}
This result shows that at very high temperature of the pion source
the distribution of
pions  will tend to become chaotic if $\theta _{k}$ is very small. 
This will
happen when the kinetic energies of the emitted pions are much larger 
than the corresponding coupling to the nucleon field,
$\omega _{k} \gg g_{k}(Y,\vec{B})$.

\newpage
\section{Conclusions}

In this paper we have proposed an intensity-dependent pion-nucleon
coupling Hamiltonian with $SU(1,1)$  dynamical symmetry. 
This Hamiltonian, within a multiparticle-production model of the AABS type
in which the $k$ - mode pion field is represented by the thermal-density 
operator, naturally explains the appearance of the NB
multiplicity distribution for pions in impact-parameter space.The shape
parameter of the NB distribution is related to the vacuum expectation 
value of the Hamiltonian.

The Wr\' oblewski - type relation (1) has been obtained with the coefficient $A$
that is energy dependent  and smaller than unity if the vacuum expectation
value of the Hamiltonian is larger than "zero-point energy" 
corresponding to $\lambda = 1/2$.

For $T \neq 0$, we have found a pion-generating function that may
be used for obtaining all  higher-order  moments of the pion field.

In our model, the $k$ modes of the pion field are statistically independent 
and are therefore described by the factorized thermal-density operator. 
Correlations
between different $k$ modes are absent and, at this stage, our model 
cannot describe the emission of resonances. However, this can be remedied
by adding  a mode-mode interacting part to the Hamiltonian  
$H(Y,\vec{B})$ [18].  We hope
to treat this case elsewhere.

\vspace{1cm}

{ \large \bf Acknowledgement }

This work was supported by the Ministry of Science of the Republic
of Croatia under Contract No.1-03-212.

\end{document}